\documentclass[aps,prapplied,twocolumn,superscriptaddress,notitlepage,nofootinbib]{revtex4-1}

\usepackage[utf8]{inputenc}
\usepackage{graphicx}
\usepackage{upgreek}
\usepackage{breakcites}
\usepackage[linktoc=page,colorlinks,urlcolor=blue,citecolor=blue,linkcolor=blue]{hyperref}
\usepackage{amssymb,amsmath}
\usepackage{graphicx}
\usepackage{natbib}
\usepackage{mathrsfs}
\usepackage{overpic}
\usepackage{wrapfig}
\usepackage{xcolor}
\usepackage{textcomp}
\usepackage{comment}
\usepackage{multirow}
\usepackage{lineno}
\usepackage{gensymb}
\usepackage{color}
\usepackage{hyperref}
\usepackage[letterpaper,textwidth=7in,top=.75in,bottom=.75in]{geometry}
\newcommand{\huPhys}{Department of Physics, Harvard University, Cambridge, MA 02138, USA}
\newcommand{\AFF}{America's Frontier Fund, Arlington, VA 22202, USA}
\newcommand{\Lam}{Lam Research Corporation, Fremont, CA 94538, USA}
\newcommand{\sEE}{Department of Electrical Engineering, Stanford University, Stanford, CA 94305, USA}
\newcommand{\sCS}{Department of Computer Science, Stanford University, Stanford, CA 94305, USA}
\newcommand{\sSRLS}{Stanford Synchrotron Radiation Lightsource, Menlo Park, CA 94025, USA}
\newcommand{\sMSE}{Department of Materials Science and Engineering, Stanford University, Stanford, CA 94305, USA}
\newcommand{\hbs}{Harvard Business School, Boston, MA 02163, USA}
\newcommand{\ucbEE}{Department of Electrical Engineering and Computer Sciences, University of California Berkeley, Berkeley, CA 94720, USA}

\begin{document}
%\linenumbers

\title{Innovating at Speed and at Scale: 
A Next Generation Infrastructure for Accelerating Semiconductor Technologies}
\date{\today}

\author{Richard A. Gottscho} 
\affiliation{\Lam}

\author{Edlyn V. Levine} 
%\email[Corresponding Author: ] {edlynlevine@aff.org}
\affiliation{\AFF} 
\affiliation{\huPhys} 

\author{Tsu-Jae King Liu} 
\affiliation{\ucbEE} 

\author{Paul C. McIntyre} 
\affiliation{\sMSE} 
\affiliation{\sSRLS} 

\author{Subhasish Mitra} 
\affiliation{\sEE}
\affiliation{\sCS}

\author{Boris Murmann} 
\affiliation{\sEE} 

\author{Jan M. Rabaey}
\affiliation{\ucbEE} 

\author{Sayeef Salahuddin}
\affiliation{\ucbEE} 

\author{Willy C. Shih} 
\affiliation{\hbs}

\author{H.-S. Philip Wong} 
\affiliation{\sEE}

\begin{abstract}
Semiconductor innovation drives improvements to technologies that are critical to modern society. The country that successfully accelerates semiconductor innovation is positioned to lead future semiconductor-driven industries and benefit from the resulting economic growth. It is our view that a next generation infrastructure is necessary to accelerate and enhance semiconductor innovation in the U.S. In this paper, we propose such an advanced infrastructure composed of a national network of facilities with enhancements in technology and business models. These enhancements enable application-driven and challenge-based research and development, and ensure that facilities are accessible and sustainable.  The main tenets are: a challenge-driven operational model, a next-generation infrastructure to serve that operational model, technology innovations needed for advanced facilities to speed up learning cycles, and innovative cost-effective business models for sustainability. Ultimately, the expected outcomes of such a participatory, scalable, and sustainable nation-level advanced infrastructure will have tremendous impact on government, industry, and academia alike.
\end{abstract}

\maketitle
\section{Introduction}
Semiconductor technologies are foundational to all modern digital economies. As the U.S. House-passed America COMPETES act heads for reconciliation with the Senate, proposals for supporting the rejuvenation of domestic semiconductor manufacturing are arriving from multiple quarters. All of the published proposals we have seen so far advocate spending more on research, training the workforce for the future, and leveraging the infrastructure that the writers either already have in place or hope to build. But a question we should ask is what kind of next-generation research and development (R$\&$D) infrastructure will be needed to meet future challenges in the face of the technological and economic obstacles that lie ahead? Equally important, what business model will ensure sustainability of these investments over the many years it will take to regain that leadership? That is what we hope to address here.

In recent years, the cost and time associated with the introduction and development of new  semiconductor technologies have drastically increased, and in many cases are prohibitive \cite{nsfFoundry, nsfCMOSX}. Growth in complexity of semiconductor devices and systems are a driving factor to these expense and delay increases, and has led to challenges in design robustness \cite{nsfEDA}. For the U.S. to remain a technology leader, it is imperative to accelerate the pace of innovation and expand the number of innovators that have access to the relevant device technology, design, and prototyping capabilities to innovate at speed and at scale. 
 
Meaningful innovation must be proven relevant to and integrated with commercial state-of-the-art devices before emerging semiconductor technologies can transition to market applications. To be both relevant and sustainable, novel approaches are needed for both the infrastructure used to conduct research and commercialization activities, and for the business models to operate and sustain that infrastructure. 

\section{Mission-Driven Infrastructure}
\begin{figure}[h]
\centering
\includegraphics[width=\columnwidth]{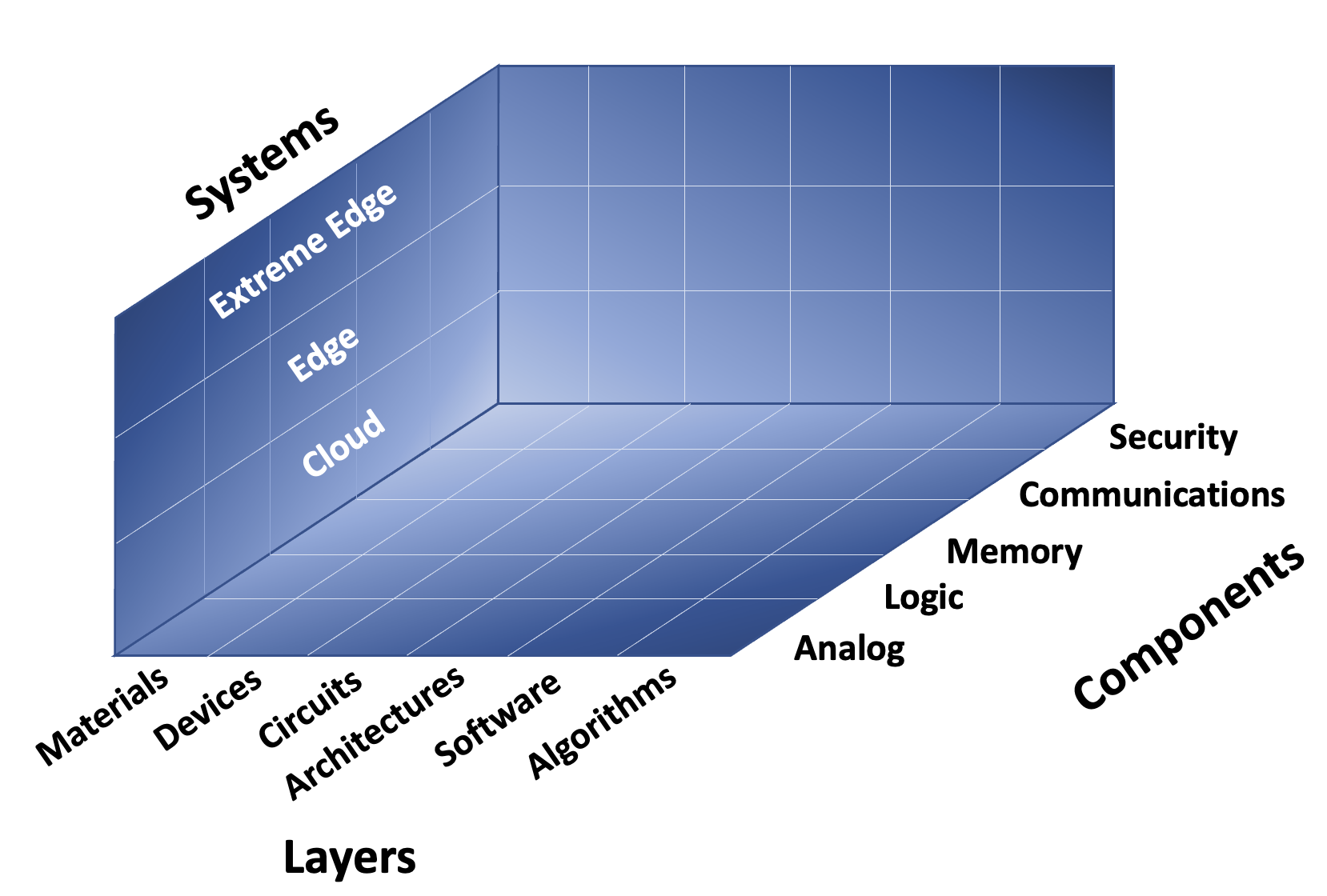}
\caption{The semiconductor and information technology innovation space. A network of facilities organized to achieve a moonshot covers sufficiently large sub-spaces to provide meaningful R$\&$D environments.}
\label{SemiEco}
\end{figure}

Semiconductor R$\&$D infrastructure has diverse requirements that depend heavily upon the specific challenges to be addressed and the part of the semiconductor ecosystem that is engaged in solving these challenges. Figure \ref{SemiEco} depicts the various axes that define the relevant ecosystem. 

The first axis follows the traditional stack model of design, with increasing abstraction from materials and devices, to architectures, systems, and applications. The second axis follows five components, which together form the building blocks of any information system: analog, digital, memory, communication, and security \cite{src}. A final axis concerns the application spaces and the associated systems, namely cloud, edge, and extreme edge \cite{bryant2020nanotechnology}. Indeed, the future requirements of the cloud (compute and data centers) are vastly different from the edge (mobiles, internet routers, edge computers, 5+G), and even further removed from the extreme edge (IoT, industrial internet, internet of actions, robots and autonomous systems). 

The axes in Figure \ref{SemiEco} represent the enormous space of diverse information technology challenges. This space is too large to be addressed by a single R$\&$D facility that covers only a subdomain. Research centers built to support the development of singular technologies such as advanced CMOS, 3D integration, storage, or analog/mixed-signal are examples of infrastructure limited to subdomains. Focused, disconnected infrastructure is the de-facto situation today for facilities at universities, company R$\&$D facilities, and national labs. This approach to infrastructure is not scalable, participatory, or sustainable within the evolving research enterprise required by the advancing complexity of information systems. 

Instead, we propose that the next generation of semiconductor R$\&$D infrastructure be designed and built around a set of moonshot goals \cite{OSTP}. Moonshots should be chosen to cover a substantial fraction of the space of Figure \ref{SemiEco}, providing context to the advances in each individual element. Such infrastructure should contain sufficient flexibility and adaptability to allow for exploration of a variety of new approaches both within and outside a moonshot program. This approach embraces co-design or co-optimization for which innovative solutions are created by considering multiple axes or aspects of a problem simultaneously. 
  
We offer two examples to illustrate the potential scope, intent and scale of a moonshot vision:
\newline
\newline
\textbf{Moonshot 1: A Virtual Model of the World} Many companies are investing in developing a virtual model of the physical world. Examples include Meta’s Metaverse and NVIDIA’s Omniverse. The realization of such virtual models could have huge impacts in domains such as health care, agriculture, climate control, mobility, etc. Current technology falls far short of the required energy efficiency needed to create a functional digital copy of the world, making breakthrough advances in the basic information processing units (here called xPUs) essential \cite{src}. Historically, the energy efficiency of graphical processing units (GPUs) has doubled every two years. We envision that a 90,000X improvement in energy efficiency of xPUs over the next 25 years will be necessary to enable a scalable virtual model of the world, requiring a  2.5X increase in energy efficiency every two years.    Achieving this moonshot will require breakthroughs on many fronts, including nanometer-scale components, innovative  logic and memory integration, advances in packaging and interconnects, mixtures of photonics and electronics, and entirely new chip architectures to implement the next generation of efficient artificial intelligence algorithms.
\newline
\newline
\textbf{Moonshot 2: A Physical Portal to the Virtual World} Benefiting from the tremendous amount of information and knowledge enabled by the virtual world envisioned by Moonshot 1 requires far richer human-computer interfaces with capabilities that extend beyond what is possible today.  Augmented VR/AR headsets are an example of such a portal. Today’s VR/AR headsets weigh 500 grams (~1 pound), have 2-3 hours of battery life, and are priced at $\$$300. Within 25 years, we envision VR/AR headsets with far richer human-computer interfaces and capabilities that weigh 10 grams or less, have 16 hours of battery life, provide data protection and security, and are priced in the same range as today’s mobile phones. Achieving this moonshot will require innovation spanning multiple disciplines including  small form factor smart sensors and actuators, mixed-signal data in-sensor data processing, neuro-inspired information extraction and interpretation,  high density low-power and low-leakage data storage, efficient communication, embedded intelligence, intrinsic security, advanced small form-factor packaging, and software stacks that are sufficiently small so as not to load down the system, yet are scalable.
\newline
\newline
Other moonshots can be envisioned, such as the  realization of human brain-like functionality in a form factor the size of the human skull at the same average power budget of 20 W, or enabling radiation-hardened cloud computing in space. The examples given serve as models and are indicative of the very different requirements imposed on the next-generation infrastructure depending upon the moonshot and the application space. One thing these moonshots have in common is the need for an accelerated pace of hardware development.

\section{General Requirements for Next-Generation Infrastructure}
To increase the pace of hardware innovation, there are three common challenges that a next-generation infrastructure must resolve. These challenges are encountered when integrating new materials, processes, and devices to realize systems that meet the metrics dictated by a  moonshot: 

\begin{enumerate}
    \item The first challenge is combining established materials and processes in an innovative way to reach desired goals. Foundries are not amenable to changing their standard process flow or parameters - e.g. dopant concentration, film thickness, etc. - to implement a new device. Since most foundry facilities are operated at full capacity to maintain profitability, they cannot be used to run the necessary experiments to develop new technology.
    \item The second challenge is even more complex and involves the introduction of new materials and processes. It is almost impossible to contract a foundry to perform the steps needed for new material introduction due to concerns about contamination and tolerance of the tools. 
    \item The third challenge is ensuring co-innovation and co-optimization across the device-to-system stack. System complexity has grown across all layers of the stack and the cost of incremental improvements within a layer have become extremely high. Continuation of siloed research domains in materials, devices, architectures, and systems is consequently unsustainable.
\end{enumerate}

Next-generation infrastructure will need to be designed and built to overcome these challenges and meet the specific research and development tasks posed by a given moonshot vision. Unconventional thinking is needed to establish the necessary infrastructure requirements to ensure success. We propose the following requirements be met by future facilities:
\newline
\newline
\textbf{Enablement of Frontier Research in CMOS and Emerging Devices:} Advances in both CMOS circuits and emerging devices must be enabled by new facilities built in the U.S. The co-innovation and co-optimization of CMOS devices with emerging devices is critical to unleash the full potential of new technologies \cite{nsfCMOSX}. U.S. government research funding has mostly been targeted towards emerging devices \cite{nsfBeyond}. Although such funding should continue, it is not surprising that research activity on advanced transistors is virtually non-existent in the U.S. This puts U.S. researchers at a global disadvantage because CMOS is the common enabler of all other technologies and the technologies developed for CMOS are broadly applicable for the manufacture of new devices and systems. Revenue coming from high-performance transistor technology dwarfs that from any other technology. Therefore, it is important to ensure that prototyping facilities allow integrated research on both CMOS and other emerging devices. 
\newline
\newline
\textbf{A Hierarchy of Experimental and Prototyping Facilities:} Next-generation infrastructure should incorporate multiple facilities to address innovation required for moonshots that span a significant subdomain of Figure \ref{SemiEco}. Laboratories will be needed to support new materials research and chemical synthesis, new equipment development and process integration, and experimental pilot lines for device and package prototyping. The requirements, tool sets, and expertise needed for each of these facilities differs greatly. Prototyping facilities will additionally need to be varied to enable experimentation on a hierarchy of wafer sizes from small substrates (coupons), to 150 mm, 200 mm wafers, and finally 300 mm wafer pilot lines. Enabling experimentation on small substrates in coupon facilities is critical since it allows for lower-cost, flexibility, and broad access to rapid experimentation for the research community. This is in contrast to 300-mm wafer pilot lines which are necessarily more rigid and not economically feasible for new materials or process development. Different process flows enabling memory, logic, and analog device prototyping must also be included and connected with advanced packaging capabilities. 
\newline
\newline
\textbf{Networking and Integration of Facilities:} The facilities constituting next generation infrastructure should be networked and integrated allowing for easy access to research teams, transfer of wafers, and shared data flows for developments to transition from device to systems level and vise-versa. Existing university laboratories, national laboratories, and commercial R$\&$D facilities should be incorporated within this network. Next generation research infrastructure must additionally be connected with commercial manufacturers to enable transition of new technologies to volume production. Standards for common technology platforms such as transistor elements, memory units, and parameters for heterogeneous integration, should be established in partnership with industry whenever possible. These standards will be important for cases in which novel systems can be built on top of existing silicon infrastructure. However, next generation infrastructure must also enable the many innovative nanomaterial, nanofabrication, and nanodevice concepts that, by nature of their novelty, do not conform to any standards.
\newline
\newline
\textbf{Enhanced Virtual Process Infrastructure:} Virtual models, often referred to as digital twins, of the tools and process flows used to prototype, scale, and manufacture new devices will drastically increase cycles of learning by enhancing empirical understanding of the introduction of new materials and process alterations. Digital twins of the process flows should be created to parallel the new physical infrastructure. These digital twins will be informed by data collected from across all facilities within the network to ensure a statistically significant data volume. Such data currently exist only within company proprietary firewalls. Open access to data generated in networked facilities will accelerate the development of virtual models. Continuous infusion of new data from ongoing experiments and prototyping will further enable digital twins of the manufacturing processes to learn and improve. Ultimately, the digital twins that are envisioned will enable researchers to simulate their innovation integrated with CMOS across all process steps to predict feasibility or improve yield upon implementation. Researcher access to the prototyping facilities is then essential to physically prove out concepts that have been modeled in the virtual world and to provide further data, thereby enabling a virtuous cycle of learning.
\newline
\newline
\textbf{Enhanced Design Tools and Flows:} Current hardware innovation severely lags the pace of software development. Long design cycles are necessary for a new device to evolve from concept to adoption in next-generation systems. To short-circuit this long dependency chain,  we envision the creation of digital twins of devices and compute systems that enable virtual and physical prototypes to co-evolve, with some parts existing in the virtual and others in the physical world. Digital twins of devices and systems allow for the injection of envisioned devices in virtual form into actual compute systems, and, vice-versa, the construction of full-scale, in-silico prototypes with newly developed devices. Achieving these digital twins requires the creation of design tools that are integrated closely with the physical and digital twin process infrastructure, and enable the early exploration and meaningful comparison between design alternatives. These tools should cover the following grounds: design space exploration and co-design across the stack, beyond traditional chip design to full system design, complete system design verification, design for  robustness in light of innovative computational paradigms,  integrity and security,  and co-creation with software.
\newline
\newline
\textbf{Cost Effective, Sustainable, and Accessible:} Next-generation infrastructure must be both cost-effective and accessible to researchers across many organizations including startups, universities, and mid- to large-scale companies. The hierarchy of facilities will require various business models to ensure sustainability. Common to all the facilities is the need for dedicated teams of cross-functional engineers to support new designs, software, controls, integration, and data flows of this next generation infrastructure. Similar to user facilities at the national labs, this permanent staff will be responsible for creating an accessible, world class user program, technique development, and research programs in collaboration with user groups. This permanent staff cannot be assignees from companies as their allegiance lies with the company where they are employed. As of today, this type of environment does not exist in the U.S. for semiconductor research. 
\newline
\newline
Next generation infrastructure must enable rapid innovation by increasing cycles of learning across the development process: design, prototyping, scaling, and production. This is possible by creating the envisioned national network of semiconductor materials laboratories, equipment development facilities, and hierarchy of prototyping lines along with their digital twins. By leveraging small diameter substrates at the onset, multiple technology vectors can be explored rapidly to support development of technologies needed to achieve the moonshots, without the fear of contamination in large wafer facilities. Digital twins of manufacturing processes and of the computational system under development further accelerate rates of learning in an ever upward spiral of innovation.

\section{Technology Innovations to Enable Next-Generation Infrastructure }

Many of the aforementioned requirements are possible to achieve with existing technology. However, some technical advances are needed to truly realize what is envisioned. 

The equipment for next generation infrastructure will need to be carefully chosen and may need to be custom built. Existing toolsets, such as 300 mm wafer processing tools, are designed for high-volume manufacturing (HVM) rather than high-throughput, iterative research. These tools are optimized for a very narrow window of operating conditions necessary to achieve the high number of wafers/hour, process uniformity, and yield needed for HVM. There are many processing tools in a 300 mm fabrication facility that each support only a single (baseline) process. Thus, even though a 300 mm facility may be intended for prototyping, fear of a tool breaking down can lead to fab-like restrictions on its operation, limiting researchers exploring new materials and process conditions. Building custom tool sets for research purposes, including tools for coupon prototyping lines, is the likely solution. Alternatively, tool manufacturers could specify a wider range of operating conditions that are allowed, while guaranteeing that the baseline process can be easily restored. 

Rapid prototyping is possible to achieve using coupon facilities; however, graduating a successful coupon prototype to a wafer prototype line as envisioned in the hierarchy of facilities is non-trivial. Doing so is contingent on developing a means to scale results from a coupon to a full wafter. This scaling problem is extraordinarily complex. Ideally, we would be able to push a button and the results from rapid coupon prototyping would be scaled and transformed to a wafer-scale design that can then be fabricated. This is not possible today and research advances are needed to achieve this goal. 

Fortunately, the same technologies that will underlie the development of  a Virtual Model of the World as articulated in Moonshot 1 and the Physical Portal into the Virtual World described by Moonshot 2 can be deployed to address the coupon-to-wafer scaling challenge. Moonshot 1 necessarily includes the digital twin of the manufacturing process envisioned in the enhanced virtual process infrastructure. Digital twins of process flows across multiple wafer sizes that learn through the continuous collection of data, leveraging of machine learning, and combination with existing physics-based models of materials properties and reactivities, have the potential to effectively determine a solution from the high-dimensional parameter space of the coupon-to-wafer problem.

Advances in high throughput metrology that provide rapid, high quality data will further enhance the learning available with virtual models. Rapid prototyping depends on rapid cycles of learning, which are in turn limited by the availability of quality metrology data. For example, in learning how to fabricate a new device, the transmission electron micrograph (TEM) is considered to be the most accurate available data. Unlike data generated by self-driving automobiles which employ inexpensive sensors and large fleets of automobiles to drive around and collect data, a TEM is complex, expensive, and laborious. Emphasis should be placed on high throughput, non-destructive, real-time metrologies to accelerate learning cycles, lower costs, and feed the digital twins.

Connecting facilities with different capabilities will require solving challenges with coupon and wafer flows that go outside one laboratory and into another. This requires carefully defining the conditions of the wafer and controlling the handling and shipping environment from the time the wafer is processed to the time it is ready for the next process. Depending on the materials and device sensitivities, different protocols and levels of control will be required. These protocols and control limits will need to be included in any new material development. Standard methodologies and metrology references will be essential. To enable rapid prototype development between facilities, it will be important to establish logistics that minimize delays, especially for queue time sensitive samples.

\section{Business Model Innovations to Enable Sustainable Facilities}

The infrastructure described here is more than just a collection of facilities. It will need funding to cover capital, operational, and research costs, but it also needs to attract a supporting ecosystem of suppliers that provides engineering and design services, and an active and robust network of engineers and innovators. The cost of running such an advanced semiconductor R$\&$D environment will not be cheap. IMEC is a 640M Euro per year operation with 4,500 employees; the NYCreates Albany facility is a $\$$300M per year operation. While government grants can provide start-up funding, what’s needed are business models that enable sustainable operation of facilities for many decades. 

The U.S. government should consider a model that encourages a competition of ideas. One that was successfully used by the National Science Foundation (NSF) in the National Strategic Computing Initiative and precursor programs was a competition to build advanced high-performance computing centers \cite{nsfNSCI}. Competitive proposals for new facilities not only had to meet technical specifications, but also had to be accompanied with business plans and initial financing contributions from partners that demonstrated long term sustainability. Additionally, because there will need to be a collection of facilities that address the broad spectrum of activities described above, the U.S. government may benefit from having overarching program management to provide coordination.

The government could award funding to build multiple advanced semiconductor infrastructure facilities – both on the strength of technical proposals and the strength of commitments from industrial, academic, and other partners. By encouraging contributions from teams of organizations with diverse skills and points of view, we can encourage a portfolio of approaches across centers to the difficult technological challenges ahead. Contributions might be in the form of funding or the commitments of resources, and the magnitude and length of commitments can be a positive factor in the awarding of centers.

The users of the facilities will likewise be granted access on a competitive basis. A significant advantage of such “bottom up” mechanisms for assigning time on shared resources is that they facilitate a wide variety of high risk/high reward research efforts to occur in parallel, although they require developing and sustaining a cadre of highly professional staff to support varied efforts that push the state of the art and, often, to develop new methods. This approach has been central to the success of the U.S. Department of Energy’s scientific user facilities that have fostered ground-breaking research advances across multiple science and engineering disciplines for many decades. 

As the government considers awarding subsidies to semiconductor manufacturers, it should consider designating some portion of those funds to purchase foundry capacity and foundry resources for exploratory research to support these proposed infrastructure centers. It could then award this capacity on a competitive basis to researchers who seek to demonstrate new device or process technologies, facilitating a critical process for scale-up of promising new technologies that would otherwise have a great deal of difficulty in accessing such capabilities. 

Having a customer for the next-generation semiconductor infrastructure centers will ensure commercial market discipline. NASA has used this model successfully over many decades to fuel advances in aeronautical research. With its Aircraft Energy Efficiency programs \cite{nasaE3} and Commercial Orbital Transportation Services (COTS) program \cite{nasaCOTS}, it ensured that there was a market pull for the technological developments that it funded and underwrote a lot of risk. This approach also encourages market competitive operations rather than subsidy-hungry facilities and avoids many of the concerns regarding IP rights when accepting government funding. 

A streamlined approach to rapid proposal development and awarding is needed to encourage private-sector participation to build new infrastructure. Incorporation of national and university laboratories into the network of facilities will provide additional incentive to private sector companies to participate by providing access to capabilities, such as advanced metrology, that they do not have on their own.

\section{Conclusion}

Semiconductors are a foundational technology that underpin a majority of technological advances for the betterment of society: from economic development and national security to combating climate change and improving medicine and healthcare. The technical community has a responsibility to formulate a research agenda for accelerating the pace of innovation that includes a plan for the R$\&$D infrastructure necessary to fulfill the high expectations of society. The research agenda advocated here is mission-oriented and challenge-driven, aiming to cover the relevant subdomains of the  semiconductor innovation space that is spanned by the three axes of systems, layers of abstraction, and component technologies. 

It is essential that the next generation R$\&$D infrastructure built to meet this research agenda is scalable, participatory, and sustainable so as to broaden the base of innovation and lower the access barrier for technology translation to commercial reality. Here, we have proposed technology and business model enhancements for facilities that enable application-driven and challenge-based innovation founded on a competition of ideas. While this infrastructure is anchored in one nation, it should have participation from technology leaders around the world for the simple reason that only the best ideas deserve focused attention. By inviting partnership with other nations that complement each other, this national infrastructure can serve to accelerate the U.S. as a global leader in R$\&$D and enable international collaboration to reach for the common good. 

\bibliographystyle{ieeetr}
\bibliography{refs}

\end{document}